\documentclass[11pt]{article}
\usepackage{amsmath,amsthm,amsfonts,amssymb, graphicx,mathabx,epsfig}
\usepackage{bbm}
\usepackage{bm}
\usepackage{authblk}
\usepackage{setspace}
\usepackage[usenames]{color}
\usepackage[active]{srcltx}

\usepackage{graphicx,color}
\usepackage{footmisc}
\usepackage{amsmath}
\usepackage{amsfonts}
\usepackage{amssymb}
\usepackage{bbm}
\usepackage{epstopdf}
\usepackage{color}
\usepackage{dsfont}
\numberwithin{equation}{section}
\usepackage{tikz}
\usepackage{pgfplots}
\usepackage{subfig}
\usetikzlibrary{fit}
\usetikzlibrary{shapes.geometric}
\usetikzlibrary{decorations.pathmorphing}
\usepackage{slashed}
\usepackage{color}
\usepackage[paper=a4paper,left=30mm,right=20mm,top=25mm,bottom=30mm]{geometry}
    \newenvironment{dedication}
        {\begin{quotation}\begin{center}\begin{em}}
        {\par\end{em}\end{center}\end{quotation}}

\renewcommand{\d}{\mathrm{d}}

\newcommand{\e}{\varepsilon}
\newcommand{\dbar}{\kern-.1em{\raise.8ex\hbox{ -}}\kern-.6em{d}}

\def\?{\marginpar{not sure}}

\newcommand{\comment}[1]{}

\newtheorem{thm}{Theorem}[section]

\newtheorem{prop}[thm]{Proposition}

\newtheorem{rem}[thm]{Remark}

\def \be{\begin{equation}}
\def \ee{\end{equation}}
\def \ben{\begin{equation*}}
\def \een{\end{equation*}}
\def \bea{\begin{eqnarray}}
\def \eea{\end{eqnarray}}

\def\qed{\hfill\raise1pt\hbox{\vrule height5pt width5pt depth0pt}}
\def\nn{\nonumber}

\def\io{\infty}

\def\Tr{\mathrm{Tr}}

\def\L{\Lambda}
\def\l{\lambda}
\def\r{\rho}
\def\s{\sigma}

\def\b{\beta}
\def\d{\delta}
\def\D{\Delta}
\def\m{\mu}

\def\e{\varepsilon}

\definecolor{light}{gray}{.75}

 \let\b=\beta    \let\d=\delta \let\e=\varepsilon
      \let\k=\kappa \let\l=\lambda
\let\m=\mu    \let\n=\nu         \let\p=\pi    \let\r=\rho
\let\s=\sigma     
   \let\o=\omega
 \let\D=\Delta  \let\L=\Lambda

\newcommand{\pp}{{\bf p}}

\newcommand{\xx}{{\bf x}}
\newcommand{\yy}{{\bf y}}

\def\nn{\nonumber}

\def\\{\hfill\break}
\def\={:=}
\let\io=\infty
\let\0=\noindent

\def\tende#1{\,\vtop{\ialign{##\crcr\rightarrowfill\crcr\noalign{\kern-1pt
    \nointerlineskip} \hskip3.pt${\scriptstyle #1}$\hskip3.pt\crcr}}\,}
\def\otto{\,{\kern-1.truept\leftarrow\kern-5.truept\to\kern-1.truept}\,}

\def\to{\rightarrow}

\def\qed{\hfill\raise1pt\hbox{\vrule height5pt width5pt depth0pt}}

\def\V#1{{\bf#1}}
\def\be{\begin{equation}}
\def\ee{\end{equation}}
\def\bea{\begin{eqnarray}}
\def\eea{\end{eqnarray}}
\def\nn{\nonumber}
\def\pref#1{(\ref{#1})}

\def\Tr{\mathrm{Tr}}

\theoremstyle{plain}

\theoremstyle{definition}

\begin{document}
\title{Canonical Drude weight for non-integrable quantum spin chains}
\author[1]{Vieri Mastropietro}
\affil[1]{University of Milano, Department of Mathematics ``F. Enriquez'', Via C. Saldini 50, 20133 Milano, Italy}
\author[2]{Marcello Porta}
\affil[2]{Eberhard Karls Universit\"at T\"ubingen, Department of Mathematics, Auf der Morgenstelle 10, 72076 T\"ubingen, Germany}
\maketitle
\begin{dedication}
{To J\"urg Fr\"ohlich, Tom Spencer and Herbert Spohn, on their 70th birthday.}
\vspace{1cm}
\end{dedication}
\begin{abstract} 
The Drude weight is a central quantity for the transport properties of quantum spin chains. The canonical definition of Drude weight is directly related to Kubo formula of conductivity. However, the difficulty in the evaluation of such expression has led to several alternative formulations, accessible to different methods. In particular, the Euclidean, or imaginary-time, Drude weight can be studied via rigorous renormalization group. As a result, in the past years several universality results have been proven for such quantity at zero temperature; remarkably the proof works for
 both for integrable and non-integrable quantum spin chains. Here we establish the equivalence of Euclidean and canonical Drude weights at zero temperature. Our proof is based on rigorous renormalization group methods, Ward identities, and complex analytic ideas.
\end{abstract}
\section{Introduction}
Linear response theory provides a convenient framework for the study of the transport properties of condensed matter systems out of equilibrium. In this paper, we shall discuss {\it non-integrable quantum spin chains}. A paradigmatic model we consider is the perturbed XXZ chain. In particular, we shall focus on its {\it Drude weight}. This quantity can be used to distinguish between metallic and insulating phases: a nonzero Drude weight corresponds to an {\it infinite conductivity}, that is perfect metallic behavior. 

For {\it integrable systems}, the zero temperature Drude weight can be typically computed from the exact solution. Let us consider the XXZ chain. Its spectrum can be computed via the Bethe ansatz \cite{YY}. The zero temperature Drude weight can be explicitly evaluated: it is nonvanishing in the gapless phase, and displays a nontrivial dependence on the model parameters. Nevertheless, at {\it zero temperature} a form of universality is believed to occur, as first predicted by Haldane \cite{Ha}: the Drude weight is expected to be related to the critical exponents and to other thermodynamic quantities, by exact, model independent relations. Such relations can be explicitely checked in the XXZ chain, thanks to the exact solution, and are conjectured to hold true also in presence of integrability breaking perturbations. 

On the contrary, at {\it positive temperature} Zotos {\it et al.} \cite{Z1, Z2} conjectured that the Drude weight depends dramatically on integrability: a {\it vanishing} Drude weight is expected to occur for non-integrable models, while a {\it finite} Drude weight should arise for integrable systems. Notice that the computation of positive temperature transport coefficients is much more involved than at zero temperature; at the moment, there is no conclusive agreement on the final value of the Drude weight \cite{ASP}. Nevertheless, in the case of the XXZ chain, a non vanishing lower bound has been proposed on the basis of the {\it Mazur bound} \cite{Maz}, see \cite{IP1,IP2, Z1, Z2}. Inspired by the above conjecture, a criterion to distinguish between interacting and non-interacting integrable systems has been recently proposed by Spohn \cite{Sp}.

Due to the sharpness of these striking claims, rigorous results for non-integrable models are particularly needed. 
A powerful way to deal with perturbations of integrable systems is provided by the {\it renormalization group} (RG). In this context, the rigorous version of the RG is especially useful: it allows to systematically take into account {\it irrelevant terms} in the RG sense, typically neglected in formal RG implementations, which encode the difference between integrable and non-integrable models. These methods have been particularly successful in proving universality at equilibrium, starting from the work of Pinson and Spencer \cite{PS, S} on the universality of the critical exponents of perturbed $2d$ Ising models (see also \cite{GGM}).

Concerning quantum spin chains, in a series of papers Benfatto and Mastropietro \cite{BM, BM1, BM2, BM3} 
constructed the Euclidean correlation functions of a class of interacting models, including the XXZ quantum spin chain in presence of non-integrable perturbations. In particular, the zero temperature Euclidean Drude weight, the critical exponents and the susceptibility were written in the form of a convergent renormalized expansion, dependent on all model details. Nevertheless, \cite{BM, BM1, BM2, BM3} proved that these quantities are indeed connected by the remarkable universality relations conjectured by Haldane. 

A key ingredient of the analysis is the rigorous comparison with an emergent relativistic quantum field theory (QFT), with fine tuned parameters depending on all lattice details, up to finite corrections. The advantage of this procedure is that the correlations of the emergent QFT can be explicitly computed, combining Schwinger-Dyson equations with chiral Ward identities (WIs). These WIs turn out to be {\it anomalous}: nevertheless, the anomaly satisfies an Adler-Bardeen-type theorem, ruling out all interaction corrections beyond the first.
The connection between the nonrenormalization of the anomalies and the universality of transport coefficients in $1d$ quantum systems was also pointed out by Alekseev, Cheianov and Fr\"ohlich in \cite{F}.
Notice that the irrelevant terms, for instance produced by the nonlinearity of the energy bands, play a crucial role for universality: they produce finite multiplicative and additive renormalizations to the transport coefficients. Their values turn out to be severely constrained by lattice conservations laws, via lattice Ward identities, which allow to determine them.

The limitation of this method is that it cannot be directly used to study real-time correlations. However, it has been recently realized that real-time transport coefficients can be computed starting from the imaginary-time ones, via a rigorous version of the {\it Wick rotation} \cite{GMPhall, AMP}. This argument combines RG-type bounds with ideas from complex analysis; it is a key ingredient of the recent proofs of the universality of the Hall conductivity for interacting Hall systems \cite{GMPhall}, and of the universality of the edge charge conductance for Hall systems with single-mode edge channel \cite{AMP}.

The main new result of the present paper is a proof of the equivalence of the Euclidean Drude weight with the canonical, or real-time, Drude weight for non-integrable $1d$ quantum spin chains. Combined with the universality results of \cite{BM, BM1, BM2, BM3}, our theorem concludes the proof of the Haldane conjecture, by extending the preexisting results to real times. As it will be clear from the proof, the extension of the strategy to positive temperatures is not at all straightforward. The obstruction in considering positive temperatures is manifest in a specific point of the proof, which will be commented below. In our opinion, this extension is a challenging open problem in mathematical and condensed matter physics, on which we plan to come back in the future. 

The paper is organized as follows. In Section \ref{sec:def} we introduce the class of models we will be interested in, together with the canonical Drude weight and the susceptibility. In Section \ref{sec:alt} we recall some alternative definitions of these transport coefficients, among which the Euclidean ones. In Section \ref{sec:univ} we recall the results of \cite{BM, BM1, BM2, BM3}. In Section \ref{sec:wick} we present the main result of this paper, together with its proof. In Section \ref{sec:concl} we summarize our results and discuss the conclusions. Finally, in Appendix \ref{app:A} we give a heuristic derivation of Kubo formula, and in Appendix \ref{app:B} we discuss the relation of canonical and thermal Drude weights.

\section{Transport coefficients of quantum spin chains}\label{sec:def}

Let $L$ be an even number, and let $\L_{L} = [-L/2, -L/2 + 1, \ldots, L/2-1, L/2] \subset \mathbb{Z}$ be a one dimensional lattice. Let $\mathbb{H} = (\mathbb{C}^{2})^{\otimes L}$, and let $S_{x}^{i} = \mathbbm{1}\otimes \mathbbm{1}\otimes \cdots \otimes \frac{\s_i}{2} \otimes \cdots \otimes \mathbbm{1}$ be an operator acting on $\mathbb{H}$, with $\s_{i}$ the $i$-th Pauli matrix, $i=1,2,3$. The Hamiltonian of the XXZ chain is defined as:
\be
\mathcal{H}_0 := - \sum_{x \in \L_{L}} (t S^1_x S^1_{x+1}+t S^2_x S^2_{x+1}+\l S^3_x S^3_{x+1})-h S^3_x+U_L \label{a1}
\ee
where $U_{L}$ fixes the boundary condition. This model is solvable by Bethe ansatz, \cite{YY}; solvability is, however, a very rigid property, typically destroyed by perturbations. For instance, consider a Hamiltonian of the form $\mathcal{H} = \mathcal{H}_{0} + \e \mathcal{V}$, where the perturbation $\mathcal{V}$ is:
\be
\mathcal{V} :=- \sum_{x\in \L_{L}}  v(x-y) S^3_x S^3_{y}\;,\label{a2}
\ee
with $v(x - y)$ of finite range. In general, we could also consider perturbations involving the products $S^{1}S^{1}$ and $S^{2}S^{2}$. These perturbed models are no longer solvable by Bethe ansatz.

In order to study the effect of the perturbation, it is convenient to map the system into a model for {\it interacting fermions}, via the Jordan-Wigner transformation. For definiteness, we choose the boundary conditions to be periodic.  Let $a^{+}_{x}$, $a^{-}_{x}$ be the fermionic creation/annihilation operators, satisfying the canonical anticommutation relations $\{ a^{+}_{x}, a^{-}_{y} \} = \d_{x,y}$, $\{ a^{+}_{x}, a^{+}_{x} \} = \{ a^{-}_{x}, a^{-}_{x} \} = 0$, with $\d_{x, y}$ the Kronecker delta. Then, it is well known that the Hamiltonian of the system can be rewritten as:
\bea
\mathcal{H}_0 &=& -\frac{t}{2}\sum_{x\in \L_{L}} (a^+_x a^-_{x+1}+a^+_{x+1}a^-_{x}) - \l\sum_{x\in \L_{L}} 
\Big(a^+_{x}a^-_{x}- \frac{1}{2}\Big)\Big(a^+_{x+1}a^-_{x+1}-\frac{1}{2}\Big)-h\sum_{x\in \L_{L}}
\Big(a^+_{x}a^-_{x}- \frac{1}{2}\Big)\nn\\
\mathcal{V} &=& -\sum_{x, y\in \L_{L}}  v(x-y) \Big(a^+_{x}a^-_{x} - \frac{1}{2}\Big)\Big(a^+_{y}a^-_{y} - \frac{1}{2}\Big)\;.
\eea
Given an operator $O$, its time evolution is $O(t) = e^{i \mathcal{H} t} O e^{-i \mathcal{H} t}$. Consider the spin density $S^{3}_{x} \equiv a^{+}_{x}a^{-}_{x} \equiv \r_{x}$. In the following, we will be particularly interested in the associated {\it spin current density} $j_{x}$, defined starting from the lattice continuity equation:
\be 
\partial_t \r_{x}(t) + d_{x} j_{x}(t) = 0
\ee
with $d_{x} f(x) := f(x) - f(x - 1)$ and
with
\be
j_x = \frac{i t}{2}(a^+_{x+1} a^-_x- a^+_{x} a^-_{x+1})\;.
\ee
The finite temperature, finite volume Gibbs state associated to $\mathcal{H}$ is denoted by $\langle \cdot \rangle_{\b, L} = \Tr \cdot e^{-\b \mathcal{H}}/\mathbb{Z}_{\b, L}$, with $\mathbb{Z}_{\b, L} = \Tr\, e^{-\b \mathcal{H}}$ the partition function. We are interested in the transport properties of the quantum spin chain, in the linear response regime. Let $\hat j_{p} = \sum_{x\in \L_{L}} e^{-ipx} j_{x}$ be the Fourier transform of the current operator, for $p\in \frac{2\pi}{L}\in \mathbb{Z}$. {\it Kubo formula} gives an expression for the {\it d.c conductivity} of the system at positive temperature, $\beta^{-1}>0$. It reads:
\be\label{ss}
\s_{\b}(p) :=  \lim_{\eta\to 0^+}  \lim_{T\to\io}  \frac{1}{
\eta} H_{T,\b}(\eta, p)
\ee
where
\bea\label{eq:HT}
H_{T,\b}(\eta, p) &:=& \lim_{L\to \infty} H_{T,\b,L}(\eta,p)\nn\\
H_{T,\b,L}(\eta,p) &:=& \frac{i}{L}\Big[\int_{-T}^{0} dt\, e^{\eta t}\, \langle [ \hat j_{p}(t)\, , \hat j_{-p} ] \rangle_{\b,L} + i\langle \D  \rangle_{\b,L} \Big]
\eea
with $\D$ is the kinetic energy operator, $\D :=- \frac{t}{2} \sum_x a^+_x a^-_{x+1}+a^+_{x+1}a^-_{x} \equiv \sum_{x} \D_{x}$. Physically, the d.c. conductivity $\s_{\b}(p)$ describes the response of the system at the time $t=0$, after introducing at the time $t = -T$ an external field oscillating in space with frequency $p$, damped by an adiabatic factor $e^{\eta t}$. For completeness, in Appendix \ref{app:A} we give a formal derivation of this formula, in the linear response approximation. Similarly, the ground state response of the system is described by:
\be 
\s_{\infty}(p) :=  \lim_{\eta\to 0^+}\lim_{T\to\io} 
\lim_{\beta\to \infty}\frac{1}{\eta} H_{T, \b}(\eta, p)\;.
\ee
Let us introduce the positive and zero temperature {\it Drude weight} as:
\be\label{eq:D}
D_{\b} := \lim_{\eta \to 0^{+}} \lim_{p\to 0} \lim_ {T\to\io} H_{T,\b}(\eta, p)\;,\qquad D_\infty := \lim_{\eta \to 0^{+}} \lim_{p\to 0} \lim_{T\to\io} \lim_{\beta\to \infty} H_{T, \b}(\eta, p)\;.
\ee
A nonzero Drude weight implies an infinite conductivity, that is perfect metallic behavior. Another possibility, not studied here, is to consider the {\it a.c. conductivity}, where the external field oscillates in time with a frequency $\o$. The zero frequency limit $\o\to 0^{+}$ of the a.c. conductivity can be formally mapped into the static limit $\eta\to 0^{+}$ of the d.c. conductivity, after replacing $\o$ with the ``imaginary frequency'' $i\eta$.

The transport properties of the system can also be investigated starting from the density-density correlation function. Let $\hat \r_p = \sum_{x\in \L_{L}} e^{-ipx} \r_{x}$ be the Fourier transform of the density operator. The positive and zero temperature {\it susceptibility} is defined as:
\be\label{eq:susc}
\kappa_{\b} := \lim_{p\to 0}\lim_{\eta \to 0^{+}}\lim_{T\to\io} K_{T,\b}(\eta, p)\;,\qquad \kappa_{\infty} := \lim_{p\to 0} \lim_{\eta \to 0^{+}}\lim_{T\to\io}\lim_{\b\to \infty}  K_{T, \b}(\eta, p)\;,
\ee
where:
\bea\label{eq:KT}
K_{T, \b}(\eta, p) &:=& \lim_{L\to \infty} K_{T, \b}(\eta, p)\nn\\
K_{T,\b,L}(\eta, p) &:=& -\frac{i}{L} \int_{-T}^{0} dt\, e^{-i\eta t}\, \langle [ \hat \r_p(t)\, , \hat \r_{-p} ] \rangle_{\b,L}\;.
\eea
We shall not discuss the rigorous validity of Eqs. (\ref{ss})--(\ref{eq:KT}) starting from many-body quantum dynamics. Instead, in the present paper we shall take Eqs. (\ref{ss})--(\ref{eq:KT}) as definitions, which will be the starting point of our rigorous analysis.

\section{Alternative definitions}\label{sec:alt}

For general, non-integrable models, the Drude weight and the susceptibility defined in the previous section are very hard to compute. We shall refer to Eqs. (\ref{eq:D}), (\ref{eq:susc}) as the {\it canonical Drude weight} and the {\it canonical susceptibility}, respectively. Due to the difficulty in evaluating these quantities, different definitions have been introduced in the literature, which are expected to be more tractable in specific cases.

Let $x_{0}\in [0, \beta)$, and let $O_{\xx} = e^{\mathcal{H} x_{0}} O_{x} e^{-\mathcal{H} x_{0}}$, with $\xx = (x_{0}, x)$, be the imaginary-time evolution of the local operator $O_{x}$: $O_{\xx} \equiv O_{x}(-ix_{0})$. This definition is then extended periodically for all $x_{0}\in \mathbb{R}$. Let $p_{0}\in \frac{2\pi}{\beta}\mathbb{Z}$ be the {\it Matsubara frequency} and $p\in \frac{2\pi}{L} [0, 1, \ldots, L-1]$ the spatial momentum. We shall collect both numbers in $\pp = (p_{0}, p)$. The space-time Fourier transform of $O_{\xx}$ is:
\be
\widehat{O}_{\pp} = \int_{0}^{\beta}dx_{0} \sum_{x\in \L_{L}} e^{-i\pp\cdot \xx} O_{\xx}\;.
\ee
We define:
\bea
H^{\text{(E)}}_{\b} (\pp) &:=& \lim_{L\to \infty} H^{\text{(E)}}_{L,\b} (\pp)\;,\\
H^{\text{(E)}}_{L,\b} (\pp) &:=& -\frac{1}{L}\langle {\bf T}\, \hat j_{\pp}\,; \hat j_{-\pp}  \rangle_{\b, L} - \frac{1}{L} \langle \D\rangle_{\b, L} \equiv -\int_{0}^{\b} dx_0 \sum_{x\in\L_L}e^{-i\pp\cdot \xx} \langle j_\xx\,; j_{\bf 0}\rangle_{\b, L} - \langle \D_0\rangle_{\b,L}\nn
\eea
where {\bf T} is the fermionic time ordering and the semicolon denotes truncation, $\langle A\, ; B \rangle = \langle AB\rangle - \langle A\rangle \langle B\rangle$. The zero temperature, infinite volume {\it Euclidean Drude weight} is:
\be\label{eq:ED}
D^{\text{(E)}}_\infty := \lim_{p_0\to 0^+} \lim_{p\to 0} \lim_{\b\to\io} H^{\text{(E)}}_{L,\b}(\pp)\;.
\ee
Similarly, let us introduce:
\bea
K^{\text{(E)}}_{\b} (\pp) &:=& \lim_{L\to \infty} K^{\text{(E)}}_{L,\b} (\pp)\nn\\
K^{\text{(E)}}_{L,\b} (\pp) &:=& \frac{1}{L}\langle {\bf T}\, \hat \r_{\pp}\,; \hat \r_{-\pp}  \rangle_{\b, L} \equiv \int_{0}^{\b}
 dx_0 \sum_{x\in\L_L}e^{-i\pp\cdot \xx} \langle \r_\xx\,; \r_{\bf 0}\rangle_{\b, L}\;.
\eea
The zero temperature {\it Euclidean susceptibility} is:
\be
\kappa^{\text{(E)}}_\infty := \lim_{p\to 0}\lim_{p_0\to 0^+} \lim_{\b\to\io}  K^{\text{(E)}}_{\b} (\pp)\;.
\ee
It turns out that, in contrast to the real-time definitions introduced in the previous section, the Euclidean transport coefficients can be conveniently studied via quantum field theory techniques.

For integrable systems, a different definition of Drude weight has been proposed, given by the following expression for finite volume and at zero temperature \cite{K, Z2}:
\be
D^{\text{(B)}}_{\infty,L} := \frac{1}{L} \frac{\partial^2 E_0(\phi)}{ \partial \phi^2}\Big|_{\phi=0}\;,
\ee
where $E_{0}(\phi)$ is the ground-state energy of the Hamiltonian of the system after a local gauge transformation:
\be 
\mathcal{H}(\phi) := - \frac{t}{2}\sum_x (e^{i\phi}a^+_x a^-_{x+1}+e^{-i\phi}a^+_{x+1}a^-_{x}) -  \l\sum_{x} \Big(a^+_{x}a^-_{x} - \frac{1}{2}\Big)\Big(a^+_{x+1}a^-_{x+1} - \frac{1}{2}\Big)\;.
\ee
Also, one can define the susceptibility as:
\be
\kappa^{\text{(B)}}_{\infty,L} := \frac{1}{L} \frac{\partial^2 E_0(h)}{\partial h^2}\Big|_{h=0}
\ee
with $E_0(h)$ the ground state energy of: 
\be
\mathcal{H}(h) := - \frac{t}{2}\sum_x (a^+_x a^-_{x+1}+a^+_{x+1}a^-_{x}) -  \l\sum_{x} \Big(a^+_{x}a^-_{x}- \frac{1}{2}\Big)\Big(a^+_{x+1}a^-_{x+1}-  \frac{1}{2}\Big) + h\sum_{x} \Big(a^+_{x}a^-_{x}-\frac{1}{2}\Big)\;.
\ee
Both $D^{\text{(B)}}_{\infty,L}$, $\kappa^{\text{(B)}}_{\infty,L}$ can be computed explicitely via Bethe ansatz \cite{YY}. One finds that, as $L\to \infty$ and setting $\cos \m= - \l/t$:
\be\label{eq:HH}
D^{\text{(B)}}_{\infty}= \frac{\pi\sin\m}{2\m (\pi-\m) }\;,\qquad \k^{\text{(B)}}_{\infty} = \frac{\m}{2\pi(\pi-\m)\sin\m }\;.
\ee
We refer the reader to \cite{Ha} for the expression of $\kappa^{\text{(B)}}_{\infty}$ (there, $v_{N} = 1/\pi\kappa^{\text{(B)}}_{\infty}$), and to \cite{GS} for the expression of $D^{\text{(B)}}_{\infty}$. It turns out that the velocity of the charge carriers can be expressed in terms of the parameters of the model as $v=\frac{\pi}{\m}\sin\m$ \cite{Ha}. Interestingly, Eqs. (\ref{eq:HH}) imply the following relation:
\be\label{kk}
\frac{D^\text{(B)}_\infty}{\k^{\text{(B)}}_{\infty}} = v^2\;.
\ee
Moreover, denoting by $K=\frac{\pi}{2(\p-\m)}$ the critical exponent of the $S^{3}$ spin-spin correlations found by Baxter in the XXZ model, one also has \cite{Ha}: 
\be
D^\text{(B)}_\infty= \frac{v K}{\pi}\;. \label{kkk}
\ee
All the quantities appearing in the relations (\ref{kk}), (\ref{kkk}) are nonuniversal, {\it i.e.} they depend on the model parameters in a nontrivial way. Nevertheless, Haldane has conjectured \cite{Ha} that the relations \pref{kk} and \pref{kkk} hold true for a large universality class, including the solvable XXZ quantum spin chain and its {\it non-integrable} perturbations. 

At positive temperature, another formulation of the Drude weight is:
\be
D^{\text{(Th)}}_\b := \lim_{T\to\io} \lim_{L\to\infty} \frac{1}{T} \int_0^T dt \int_0^\b d x_{0} \sum_{x = -L/2}^{L/2} \langle j_x(t-i x_{0}) j_0(0)\rangle_{\b,L}\;.
\ee
$D^{\text{(Th)}}_{\b}$ is called the {\it thermal Drude weight}. Equivalently, if the current-current correlation function reaches its limit fast enough as $t\to \infty$:
\be\label{eq:Dth}
\widetilde{D}^{\text{(Th)}}_\b := \lim_{t\to\io} \lim_{L\to\infty} \int_0^\b d x_{0}
\sum_{x = -L/2}^{L/2} \langle j_x(t-i x_{0}) j_0(0)\rangle_{\b,L}\;.
\ee
As far as we know, the identity between $\widetilde{D}^{\text{(Th)}}_\b$ or $D^{\text{(Th)}}_{\b}$
and the canonical Drude weight is unproven; see, {\it e.g.}, \S 6 of \cite{IP1}. In Appendix \ref{app:B} we prove the equivalence between the canonical Drude weight, as defined by Eq. (\ref{eq:D}), and a suitably regularized version of $\widetilde{D}^{\text{(Th)}}_{\b}$. 

The utility of $\widetilde{D}^{\text{(Th)}}_{\b}$ is that it can be bounded from below thanks to the {\it Mazur bound} \cite{Maz}; this estimate implies that $\widetilde{D}^{\text{(Th)}}_\b$ is nonzero in the XXZ model  \cite{Z1, IP1, IP2}.
Based on this observation, Spohn \cite{Sp} proposed a criterion to distinguish between non-interacting and interacting integrable systems, depending on whether the following quantity is zero or not (using the shorthand notation $\sum_{x} \langle \cdot \rangle_{\b} \equiv \lim_{L\to \infty} \sum_{x=-L/2}^{L/2} \langle \cdot \rangle_{\b, L}$): 
\be
\mathcal{L} := \int_{\mathbb{R}} dt\,  \Big[\int_0^\b dx_{0} \sum_{x}
\langle j_x(t - ix_{0}) j_0(0)\rangle_{\b}\;-\lim_{t\to\io} \int_0^\b dx_{0} \sum_{x}
\langle j_x(t - i x_{0}) j_0(0)\rangle_{\b}\Big]\;.
\ee
In the XX spin chain $\mathcal{L} = 0$, while it is expected that $\mathcal{L} \neq 0$ in the XXZ chain.

Finally, another way of computing the Drude weight is {\it dynamically}: one considers a large but finite quantum spin chain with open boundary conditions, and introduces a nontrivial chemical potential gradient. Then, a nonzero Drude weight is related to a linear increase in time of the total current flowing in the chain. This procedure is particularly convenient for numerical simulations, \cite{Ka, Bu, IN, LLMM}.

\section{Universality relations for Euclidean transport coefficients}\label{sec:univ}

The definitions of Euclidean Drude weight and susceptibility are particularly useful in the study of {\it nonsolvable} quantum spin chains. Both quantities have been computed explicitly in \cite{BM2}, \cite{BM3} (see Theorem 1.1 of 
\cite{BM2} and Theorem 1.1 of \cite{BM3}). We summarize here the main results. For convenience, in the following we shall set:
\be
j_{0,\xx} \equiv \r_{\xx}\;,\qquad j_{1,\xx} \equiv j_{\xx}\;.
\ee
\begin{thm}{\bf (Euclidean transport coefficients.)}\label{thm:1} Consider the nonsolvable quantum spin chain with Hamiltonian $\mathcal{H} = \mathcal{H}_0 + \e \mathcal{V}$, defined in Eqs. \pref{a1}, \pref{a2}. Then, for $|\l|, |\e|$ small enough, the following is true.
\begin{itemize}
\item[(i)] The Euclidean current-current correlation functions are analytic in $\l, \e$, and satisfy the following bound, for all $x_{0}, y_{0} \in [0, \beta)$, $x, y\in \L_{L}$:
\be\label{eq:xx}
| \langle {\bf T}\, j_{\m,\xx}\,; j_{\n,\yy}\rangle_{\beta,L}| \leq \frac{C}{1 + | \xx - \yy |^2}\;,
\ee
with $C$ independent of $\b, L$, and $|\cdot|$ the distance on the torus of sides $\beta, L$.
\item[(ii)] Let $\pp_{\b,L} \in \frac{2\pi}{\beta}\mathbb{Z}\times \mathbb{T}_{L}$, $\pp_{\b,L}\neq (0,0)$, with $\pp_{\b,L} \to \pp\in \mathbb{R}\times \mathbb{T}$ as $\beta, L \to \infty$, $\pp\neq (0,0)$. Then:
\bea\label{eq:JJbd}
&&\lim_{\beta, L\to \infty} \int_{0}^{\beta}dx_{0} \sum_{x\in \L_{L}} e^{-i\pp_{\b,L}\cdot \xx} \langle j_{\m,\xx}\,; j_{\n,\00}\rangle_{\beta,L} \quad \text{exists,}\nn\\&&\;\qquad \Big| \int_{0}^{\beta}dx_{0} \sum_{x\in \L_{L}} e^{-i\pp_{\b,L}\cdot \xx} \langle j_{\m,\xx}\,; j_{\n,\00}\rangle_{\beta,L} \Big| \leq C
\eea
with $C$ independent of $\b, L, \pp_{\b, L}$. Moreover,
\bea\label{eq:Dkappa}
\lim_{\b\to\io}H^{\text{(E)}}_{\b}(\pp) &=& \frac{v K}{\pi}\frac{p_0^2}{p_0^2+v^2 p^2}+R_{H}(\pp)\nn\\
\lim_{\b\to\io}K^{\text{(E)}}_{\b}(\pp)&=&\frac{K}{\pi v}\frac{v^2 p^2}{p_0^2+v^2  p^2}+R_{K}(\pp)
\eea
where $K \equiv K(\l, \e)>0$, $v\equiv v(\l, \e) >0$ are analytic in $\l,\e$ and $R_{H}$, $R_{K}$ are continuos in $\pp$, such that $R_{H}(0,0)= R_{K}(0,0)=0$. That is, the Euclidean Drude weight and the Euclidean susceptibility are given by:
\be
D^{\text{(E)}}_{\infty} = \frac{v K}{\pi}\;,\qquad \kappa^{\text{(E)}}_{\infty} = \frac{K}{v\pi}\;.
\ee

\end{itemize}  
\end{thm}
In particular, the above theorem establishes the validity of the Haldane universality relations \cite{Ha}, at zero temperature, for the Euclidean Drude weight and the Eucllidean susceptibility, in the nonintegrable case $\l\neq 0$, $\e\neq 0$. The parameter $v$ has the interpretation of dressed Fermi velocity for the interacting system. A similar result can be proven for a larger class of perturbations of the XXZ chain, involving the operators $S^{1} S^{1}$ or $S^{2} S^{2}$ \cite{M1}. 

The main technical tool behind the proof of this theorem is the rigorous renormalization group. This method can be used to construct the zero temperature, infinite volume limit of all correlation functions, and allows to prove bounds on their decay, such as Eq. (\ref{eq:xx}). Notice, however, that this bound is not even enough to prove the finiteness of the Drude weight and of the susceptibility. In order to compute these transport coefficients, one has to exploit cancellations, by using {\it Ward identities}.

 Let us briefly sketch the main steps of the proof. The starting point is a rewriting of the correlations of the interacting spin chain in terms of a Grassmann integral, which can be evaluated via a multiscale integration. Every step of integration is expressed thorugh a convergent expansion in terms of suitable running coupling constants. Convergence is based on an expansion in truncated expectations (rather than Feynman diagrams) and on {\it determinant bounds}, \cite{M2, BFM}.
 
In order to compute the transport coefficients, we compare their expansion with the one for a suitable {\it reference model}, with fine tuned bare parameters. This model describes a relativistic $1+1$ dimensional quantum field theory, whose correlations can be computed explicitely via chiral Ward identities. This comparison is possible thanks to the fact that the difference between the two models is encoded into {\it irrelevant terms} in the RG sense. It is important to stress that this does not mean that these terms can be simply neglected: they produce {\it finite corrections} to the physical quantities, such as the Drude weight and the susceptibility, which are essential for the proof of the Haldane conjecture.

An important feature of the reference model is the presence of the {\it chiral anomaly} in the Ward identities, produced by the ultraviolet momentum cutoff, which has to be introduced in order to properly define the relativistic theory. The anomalies verify an Adler-Bardeen-type nonrenormalization property, which is essential for the universality of the Haldane relations. This construction ultimately allows to write $H^{\text{(E)}}_{\infty}(\pp)$, $K^{\text{(E)}}_{\infty}(\pp)$ as the sum of a discontinuous term, whose structure is determined by the reference model and by the combination of lattice and chiral Ward identities, plus continuous contributions, depending on all details of the model, vanishing as $\pp\to (0,0)$.

The missing step in the above construction is the connection with the real-time transport coefficients, as defined in Section \ref{sec:def}. The main goal of the present paper is to fill this gap, by extending Theorem \ref{thm:1} to the {\it canonical} Drude weight and susceptibility at zero temperature.

\section{Equivalence of canonical and Euclidean Drude weights}\label{sec:wick}

In this section we shall prove the identity between the Euclidean and canonical Drude weights. Thus, we will be able to compute them, thanks to Theorem \ref{thm:1}. Technically, we shall prove the equivalence between Euclidean and canonical transport coefficients via a rigorous version of the {\it Wick rotation}, for non-integrable, interacting quantum spin chains.
\begin{thm}{\bf (Wick rotation.)}\label{thm:wick} Under the assumptions of Theorem \ref{thm:1}, the following is true. Let $\eta_{\b}$ be the closest element of $\frac{2\pi}{\beta}\cdot \mathbb{Z}$ to $\eta$, that is $|\eta - \eta_{\beta}| = \text{min}_{\eta'\in \frac{2\pi}{\beta}\cdot \mathbb{Z}} |\eta - \eta'|$. Then, the following identities hold true:
\bea\label{eq:HK}
H_{T,\beta,L}(\eta, p) &=& -\int_{0}^{\beta} dt\, e^{-i\eta_{\beta} t} \frac{1}{L}{\langle} \hat j_{p}(-it)\,;  \hat j_{-p} {\rangle}_{\beta, L} - \langle \D_{0}\rangle_{\b,L} + \mathcal{E}^{(H)}_{\b,L}(T,\eta)\nn\\
K_{T,\beta,L}(\eta, p) &=& \int_{0}^{\b} dt\, e^{-i\eta_{\b} t} \frac{1}{L}\langle \hat \r_p(-it)\,; \hat \r_{-p}\rangle_{\b, L} +\mathcal{E}^{(K)}_{\b,L}(T,\eta)
\eea
where the error terms $\mathcal{E}^{(K)}_{\b,L}(T,\eta)$, $\mathcal{E}^{(H)}_{\b,L}(T,\eta)$ satisfy the bounds:
\be
\big|\mathcal{E}^{(\sharp)}_{\b,L}(T,\eta)\big|\leq C\Big( \frac{1}{\eta^2 \beta} + e^{-\eta T}\Big)\;,\qquad \sharp = H,\, K\;,
\ee
for some $C>0$ independent of $\beta, L, \eta, T$. Moreover, the limits $\lim_{\beta \to \infty}\lim_{L\to \infty}H_{T,\beta,L}(\eta, p)$, $\lim_{\beta\to \infty}\lim_{L\to \infty} K_{T,\beta,L}(\eta, p)$ exist, and are given by:
\bea\label{eq:Hinfty}
\lim_{\beta\to \infty} \lim_{L\to \infty} H_{T,\beta,L}(\eta, p) &=& H^{\text{(E)}}_{\infty}(\eta, p) + \mathcal{E}^{(H)}_{\infty}(T,\eta)\nn\\
\lim_{\beta\to \infty} \lim_{L\to\infty} K_{T,\beta,L}(\eta, p) &=&K_{\infty}^{\text{(E)}}(\eta, p) +\mathcal{E}^{(K)}_{\infty}(T,\eta)
\eea
with $H^{\text{(E)}}_{\infty}(\eta, p)$, $K^{\text{(E)}}_{\infty}(\eta, p)$ the Euclidean Drude weight and susceptibility given by Eqs. (\ref{eq:Dkappa}), and where $|\mathcal{E}^{(\sharp)}_{\infty}(T,\eta)|\leq Ce^{-\eta T}$. Finally,
\bea\label{eq:Hfin}
\lim_{T\to \infty} \lim_{\beta\to \infty}\lim_{L\to \infty} H_{T,\beta,L}(\eta, p) &=& H^{\text{(E)}}_{\infty}(\eta, p)\nn\\
\lim_{T\to \infty} \lim_{\beta\to \infty}\lim_{L\to \infty} K_{T,\beta,L}(\eta, p) &=& K^{\text{(E)}}_{\infty}(\eta, p)\;.
\eea
\end{thm}
Therefore, combining Eqs. (\ref{eq:D}), (\ref{eq:susc}) with Eqs. (\ref{eq:Hfin}), (\ref{eq:Dkappa}), we get the announced identity between Euclidean and canonical transport coefficients:
\bea
D_\io &=& \lim_{\eta\to 0^+} \lim_{p\to 0} \lim_{T\to \infty} \lim_{\beta\to \infty}  H_{T,\beta}(\eta, p)=\lim_{\eta\to 0^+} \lim_{p\to 0} H^{\text{(E)}}_{\infty}(\eta, p)=D^{\text{(E)}}_\io\nn\\
\kappa_{\infty} &=& \lim_{p\to 0} \lim_{\eta \to 0^{+}}\lim_{T\to\io}\lim_{\b\to \infty}  K_{T, \b}(\eta, p)=\lim_{p\to 0} \lim_{\eta \to 0^{+}} K^{\text{(E)}}_{\infty}(\eta, p) = \kappa_{\infty}^{\text{(E)}}\;.
\eea
The rest of the paper is devoted to the proof of Theorem \ref{thm:wick}. In the following proposition we collect some technical results, that will be used in the proof of our main result.
\begin{prop}\label{prp:lim} Under the same assumptions of Theorem \ref{thm:1}, the following is true.
\begin{itemize}
\item[(i)] The limit of the static correlation functions $\lim_{\beta, L\to \infty} \langle a^{\e_{1}}_{x_{1}}\cdots a^{\e_{n}}_{x_{n}} \rangle_{\b, L}$ exists.
\item[(ii)] The following bound holds true:
\be\label{eq:bd}
\Big|\frac{1}{L}{\langle} \hat j_{\m, p}(z) \hat j_{\n, -p} {\rangle}_{\beta, L}\Big|\leq C\;,\qquad \text{Im}\,z\leq 0\;,
\ee

uniformly in $\beta, L, z, p$.
\item[(iii)] The limit
\be\label{eq:lim}
\lim_{\beta\to\infty} \lim_{L\to\infty}  \frac{1}{L}{\langle} [ \hat j_{\m, p}(t)\, , \hat j_{\n, -p}] {\rangle}_{\beta, L} \qquad \text{exists for all $t\in\mathbb{R}$.}
\ee
\end{itemize}
\end{prop}
%

For the moment, let us postpone the proof of Proposition \ref{prp:lim}, and let us show how Proposition \ref{prp:lim} can be used to prove Theorem \ref{thm:wick}.
\begin{proof} (of Theorem \ref{thm:wick}.) We shall prove the statement for the Drude weight. The proof of the result for the susceptibility is completely analogous, and will be omitted. Let $\pp = (\eta, p) \neq (0,0)$. Consider the time integral appearing in the definition of $H_{T, \b, L}$, Eq. (\ref{eq:HT}). We claim that:
\be\label{eq:cl1}
\int_{-T}^{0} dt\, e^{\eta t} \frac{1}{L}{\langle} [ \hat j_{p}(t)\, , \hat j_{-p}] {\rangle}_{\beta, L} = i\int_{0}^{\beta} dt\, e^{-i\eta_{\beta} t} \frac{1}{L}{\langle} \hat j_{p}(-it)\,;  \hat j_{-p} {\rangle}_{\beta, L} + \mathcal{E}^{\text{(H)}}_{\beta, L}(T,\eta)\;,
\ee
where $\eta_{\beta} \in \frac{2\pi}{\beta}\cdot \mathbb{Z}$ is such that $|\eta - \eta_{\beta}| = \text{min}_{\eta'\in \frac{2\pi}{\beta}\cdot \mathbb{Z}} |\eta - \eta'|$, and where the error term satisfies the bound $\big| \mathcal{E}^{\text{(H)}}_{\beta, L}(T,\eta) \big| \leq C( 1/(\eta^2 \beta) + e^{-\eta T})$. This claim immediately implies Eq. (\ref{eq:HK}). Let us postpone for a moment the proof of Eq. (\ref{eq:cl1}), and let us first discuss the $\beta, L\to \infty$ limit of Eq. (\ref{eq:cl1}). The existence of the limit for the left-hand side is implied by Proposition \ref{prp:lim}, item $(iii)$. The existence of the limit of the first term in the right-hand side is implied by Theorem \ref{thm:1}, item $(ii)$. Thus,
\be
\int_{-T}^{0} dt\, e^{\eta t} \lim_{\beta,L\to \infty}\frac{1}{L}{\langle} [ \hat j_{p}(t)\, , \hat j_{-p}] {\rangle}_{\beta, L} = i\lim_{\b, L\to \infty}\int_{0}^{\beta} dt\, e^{-i\eta_{\beta} t} \frac{1}{L}{\langle} \hat j_{p}(-it)\,;  \hat j_{-p} {\rangle}_{\beta, L} + \mathcal{E}^{\text{(H)}}_{\infty}(T,\eta)
\ee
with $|\mathcal{E}^{\text{(H)}}_{\infty}(T,\eta)| \leq Ce^{-\eta T}$. Also, the existence of $\lim_{\beta, L\to \infty} \langle \D_{0} \rangle_{\b, L}$ follows from Proposition \ref{prp:lim}. This proves Eq. (\ref{eq:Hinfty}). Taking also the $T\to\infty$ limit we have:
\be
\int_{-\infty}^{0} dt\, e^{\eta t} \lim_{\beta,L\to \infty}\frac{1}{L}{\langle} [ \hat j_{p}(t)\, , \hat j_{-p}] {\rangle}_{\beta, L} = i\lim_{\b, L\to \infty}\int_{0}^{\beta} dt\, e^{-i\eta_{\beta} t} \frac{1}{L}{\langle} \hat j_{p}(-it)\,;  \hat j_{-p} {\rangle}_{\beta, L}
\ee
which proves the final claim, Eq. (\ref{eq:Hfin}). Thus, we are left with proving Eq. (\ref{eq:cl1}). The proof this identity is based on a complex deformation argument, which goes as follows. We start by writing:
\bea\label{eq:W1}
\int_{-T}^{0} dt\, e^{\eta t} {\langle} [ \hat j_{p}(t) , \hat j_{-p} ] {\rangle}_{\beta, L} &=& \int_{-T}^{0} dt\, \big[ e^{\eta t} {\langle} \hat j_{p}(t) \hat j_{-p} {\rangle}_{\beta, L} - e^{\eta t} {\langle} \hat j_{-p} \hat j_{p}(t) {\rangle}_{\beta, L}\nn\\
&=& \int_{-T}^{0} dt\, \big[ e^{\eta t} {\langle} \hat j_{p}(t) \hat j_{-p} {\rangle}_{\beta, L} - e^{\eta t} {\langle} \hat j_{p}(t - i\beta) \hat j_{-p} {\rangle}_{\beta, L} \big]\;,
\eea
where in the last line we used that, by cyclicity of the trace $\Tr e^{-\beta \mathcal{H}} \hat j_{-p} \hat j_{p}(t) = \Tr \hat j_{p}(t) e^{-\beta \mathcal{H}} \hat j_{-p} = \Tr e^{-\beta \mathcal{H}} \hat j_{p}(t - i\beta) \hat j_{-p}$. Let $\eta_{\beta}$ be the closest element of $\frac{2\pi}{\beta}\cdot \mathbb{Z}$ to $\eta$: $|\eta - \eta_{\beta}| = \min_{\eta'\in \frac{2\pi}{\beta}\cdot \mathbb{Z}} |\eta - \eta'|$. We rewrite (\ref{eq:W1}) as, using that $e^{i\eta_{\beta}\beta} = 1$:
\bea\label{eq:W11a}
\int_{-T}^{0} dt\, e^{\eta t} \frac{1}{L}{\langle} [ \hat j_{p}(t) , \hat j_{-p} ] {\rangle}_{\beta, L} &=& \int_{-T}^{0} dt\, \big[ e^{\eta_{\beta} t} \frac{1}{L}{\langle} \hat j_{p}(t) \hat j_{-p} {\rangle}_{\beta, L} - e^{\eta_{\beta} (t - i\beta)} \frac{1}{L}{\langle} \hat j_{p}(t - i\beta) \hat j_{-p} {\rangle}_{\beta, L} \big]\nn\\&& + \mathcal{E}^{(1)}_{\beta, L}(T, \eta)\;.
\eea
The error term can be estimated using Eq. (\ref{eq:bd}):
\be\label{eq:E1}
\big|\mathcal{E}^{(1)}_{\beta, L}(T, \eta)\big| \leq C \int_{-T}^{0}dt\, \big|e^{\eta t} - e^{\eta_{\beta} t}\big| \leq \frac{C}{\eta^2 \beta}\;.
\ee
Then, we notice that the function $z\mapsto e^{\eta_{\beta} z} {\langle} \hat j_{p}(z) \hat j_{-p} {\rangle}_{\beta,L}$ is {\it entire} in $z\in \mathbb{C}$. Hence, by Cauchy theorem we conclude that the integral of $e^{\eta_{\beta} z} {\langle} \hat j_{p}(z) \hat j_{-p} {\rangle}_{\beta,L}$ along the boundary of the complex rectangle $(0,0)\to (0,-i\beta)\to (-T,-i\beta) \to (-T,0)$ is equal to zero. The two terms in Eq. (\ref{eq:W11a}) correspond respectively to the paths $(-T, 0)\to (0,0)$ and $(0, -i\beta) \to (-T, -i\beta)$. Thus we can write:
\bea
&&\int_{-T}^{0} dt\, \big[ e^{\eta_{\beta} t} \frac{1}{L}{\langle} \hat j_{p}(t) \hat j_{-p} {\rangle}_{\beta, L} - e^{\eta_{\beta} (t - i\beta)} \frac{1}{L}{\langle} \hat j_{p}(t - i\beta) \hat j_{-p} {\rangle}_{\beta, L} \big] \nn\\&&\qquad  =  i\int_{0}^{\beta} dt\, e^{-i\eta_{\beta} t} \frac{1}{L}{\langle} \hat j_{p}(-it) \hat j_{-p} {\rangle}_{\beta, L} + \mathcal{E}^{(2)}_{\beta, L}(T, \eta)\;,
\eea
where the new error term collects the contribution coming from the integration over the path $(-T,-i\beta)\to (-T,0)$,
\bea\label{eq:E2}
\mathcal{E}^{(2)}_{\beta, L}(T, \eta) &=& -i\int_{0}^{\beta} dt\, e^{\eta_{\beta}(-T -it)} \frac{1}{L}{\langle} \hat j_{p}(-T -it) \hat j_{-p} {\rangle}_{\beta,L}\nn\\
\big| \mathcal{E}^{(2)}_{\beta, L}(T, \eta) \big| &\leq& e^{-\eta T} \frac{1}{L}\int_{0}^{\beta}dt\, {\langle} \hat j_{p}(-it)\hat j_{-p} {\rangle}_{\beta,L}^{1/2}{\langle} \hat j_{p}(-it) \hat j_{-p} {\rangle}_{\beta,L}^{1/2}\\
&\leq& e^{-\eta T} \Big(\int_{0}^{\beta}dt\, \frac{1}{L} {\langle} \hat j_{p}(-it) \hat j_{-p} {\rangle}_{\beta,L}\Big)^{1/2}\Big(\int_{0}^{\beta}dt\, \frac{1}{L} {\langle} \hat j_{p}(-it)\hat j_{-p} {\rangle}_{\beta,L}\Big)^{1/2} \leq Ce^{-\eta T}\;.\nn
\eea
The first bound follows from the cyclicity of the trace and from the Cauchy-Schwarz inequality for $\langle \cdot \rangle_{\beta, L}$ (see the proof of Proposition \ref{prp:lim} for more details), while the second follows from the Cauchy-Schwarz inequality for the time integration. In this way, we reconstructed the Fourier transforms of the current-current correlation functions, that can be bounded thanks to Eq. (\ref{eq:JJbd}). Therefore, all in all we have:
\be
\int_{-T}^{0} dt\, e^{\eta t} \frac{1}{L}{\langle} [ \hat j_{p}(t) \,, \hat j_{-p} ] {\rangle}_{\beta, L} = i\int_{0}^{\beta} e^{-i\eta_{\beta} t} \frac{1}{L}{\langle} \hat j_{p}(-it)\,;  \hat j_{-p} {\rangle}_{\beta, L} + \sum_{i=1}^{2} \mathcal{E}^{(i)}_{\beta, L}(T, \eta)
\ee
where, by Eqs. (\ref{eq:E1}), (\ref{eq:E2}), $\big| \sum_{i=1}^{2} \mathcal{E}^{(i)}_{\beta, L}(T, \eta) \big| \leq C( 1/(\eta^2 \beta) + e^{-\eta T})$. The semicolon in the right-hand side can be inserted for free, since $\langle \hat j_{p} \rangle_{\beta, L} = 0$. This concludes the proof.
\end{proof}

\begin{rem} In Eq. (\ref{eq:cl1}), it is essential that the limit $\beta\to \infty$ is taken {\em before} the limit $\eta \to 0$. If one exchanges the limits the error term $1/(\eta^2 \beta)$ blows up as $\eta \to 0^+$. This is the main obstruction to the extension of the proof to the positive temperature case.
\end{rem}

To conclude, we are left with proving Proposition \ref{prp:lim}.

\begin{proof}(of Proposition \ref{prp:lim}.) Consider item $(i)$. The proof of the existence of the zero temperature, infinite volume limit of the configuration space correlations follows from the application of cluster expansion methods, see {\it e.g.} \cite{BFM}; we will omit the details.

Let us prove the bound on the complex times current-current correlation, item $(ii)$. This estimate can be proven using the cyclicity of the trace and the Cauchy-Schwarz inequality for $\langle \cdot \rangle_{\b, L}$. We have:
\bea
&&|{\langle} \hat j_{\m, p}(z) \hat j_{\n, -p} {\rangle}_{\beta, L}| = |{\langle} \hat j_{\m, p}(z/2) \hat j_{\n, -p}(-z/2) {\rangle}_{\beta, L}| \\&&\quad \leq  \langle \hat j_{\m, p}(i\text{Im} z/2) \hat j_{\m, -p}(-i\text{Im} z/2)\rangle_{\b, L}^{1/2}  \langle \hat j_{\n, p}(i\text{Im}z/2) \hat j_{\n, -p}(-i\text{Im}z/2) \rangle_{\b, L}  ^{1/2}\;,\nn
\eea
where we used the following identities:
\bea
\hat j_{\m, p}(z) &=& e^{iz \mathcal{H}_{L}} \hat j_{\m, p} e^{-i z\mathcal{H}_{L}} = e^{i \text{Re} z \mathcal{H}_{L}} \hat j_{\m, p}(i \text{Im} z) e^{-i \text{Re} z \mathcal{H}_{L}}\nn\\
\hat j_{\m, p}(z)^{*} &=&  e^{i \bar z \mathcal{H}_{L}} \hat j_{\m, -p} e^{-i \bar z \mathcal{H}_{L}} = e^{i \text{Re} z \mathcal{H}_{L}} \hat j_{\m, -p}(-i \text{Im} z) e^{-i \text{Re} z \mathcal{H}_{L}}\nn\\
\hat j_{\m, p}(z) \hat j_{\m, p}(z)^* &=& e^{i \text{Re} z \mathcal{H}_{L}}  \hat j_{\m, p}(i\text{Im} z)  \hat j_{\m, -p}(-i \text{Im} z) e^{-i \text{Re} z \mathcal{H}_{L}}\;.
\eea
It is important to notice that, after using Cauchy-Schwarz and the cyclicity of the trace, the real part of $z$ {\it cancels}, thus leaving us with a bound only involving Euclidean correlations. Then, in order to prove the bound in Eq. (\ref{eq:bd}) we use that, thanks to the bound (\ref{eq:xx}) for the decay of the Euclidean current-current correlation function:
\be
\frac{1}{L} \langle \hat j_{\m, p}(i\text{Im}z/2) \hat j_{\m, -p}(-i\text{Im}z/2) \rangle_{\b, L} \equiv \frac{1}{L} \langle \hat j_{\m, p}(i\text{Im}z)\,; \hat j_{\m, -p}\rangle_{\b, L} \leq \sum_{x} \frac{C}{1 + (\text{Im}z)^2+ x^2} \leq C'\;.
\ee
To get the last bound, we used that, thanks to the assumption $\text{Im}z\leq 0$, $\langle \hat j_{\m, p}(i\text{Im}z)\,; \hat j_{\m, -p}\rangle_{\b, L} \equiv \langle {\bf T}\, \hat j_{\m, p}(i\text{Im}z)\,; \hat j_{\m, -p}\rangle_{\b, L}$. This proves item $(ii)$.

Finally, let us prove the existence of the time-dependent correlations, item $(iii)$. By translation invariance, we can rewrite
\be
\frac{1}{L}{\langle} [ \hat j_{\m, p}(t)\, , \hat j_{\n, -p}] {\rangle}_{\beta, L} = \sum_{x} e^{-ipx} \langle [ j_{\m, x}(t)\, , j_{\n, 0} ] \rangle_{\beta, L}\;.
\ee
By the {\it Lieb-Robinson bound} \cite{LR, N}, we know that the sum converges uniformly in $\beta, L$: 
\be
\| [ j_{\m, x}(t)\, , j_{\n, 0} ]  \| \leq C_{M}\frac{e^{vt}}{1 + |x|^{M}}\;,\qquad \text{for all $M\geq 1$} 
\ee
with $|\cdot|$ the Euclidean distance on the periodic lattice $\L_{L}$. Therefore, to prove the existence of the limit in Eq. (\ref{eq:lim}) it is enough to show that:
\be
\lim_{\b\to\infty} \lim_{L\to\infty} \langle [ j_{\m, x}(t)\, , j_{\n, 0} ] \rangle_{\beta, L}\qquad \text{exists for all $t$.}
\ee
The proof of this last fact is based on the existence of the infinite volume dynamics, together with the existence of the zero temperature, infinite volume Gibbs state. Let us sketch the argument. Let $L \ge R$. We rewrite:
\be\label{eq:decomp}
 \big\langle j_{\m, x}(t) j_{\n, 0} \big\rangle_{\beta,L} = \big\langle j_{\m, x}^{R}(t) j_{\n, 0} \big\rangle_{\beta,L} + \big\langle \big(j_{\m, x}(t) - j_{\m, x}^{R}(t)\big) j_{\n, 0} \big\rangle_{\beta,L}
\ee
with $j^{R}_{\m, x}(t) = e^{i\mathcal{H}_{L}^{R}t} j_{\m, x} e^{-i\mathcal{H}_{L}^{R}t}$, where the Hamiltonian $\mathcal{H}_{L}^{R}$ only contains potentials with support in $\L_{R} = [-R/2, R/2]$:
\be
\mathcal{H}_{L}^{R} = \sum_{X\subset \L_{R}} \Phi^{L}_{X}\;;
\ee
$\{\Phi^{L}_{X}\}$ are the potentials appearing in the original Hamiltonian $\mathcal{H}_{L}$ (the $L$-dependence only comes from the requirement of periodic boundary condition). For $L$ large enough and $R$ fixed, the potentials in the sum are actually $L$-independent, thanks to the finite range of hopping and interaction potential. By a standard application of Lieb-Robinson bounds, one gets \cite{N}:
\be\label{eq:NOS}
\big\| j_{\m, x}^{R}(t)  - j_{\m, x}(t) \big\| \leq \sum_{y\in \L_{L}\setminus \L_{R}}\frac{C_{X,M}(t)}{1 + |x - y|^{M}}\;,\qquad \forall M\in\mathbb{N},\;
\ee
with $C_{X,M}(t)$ independent of $L$ (exponentially growing with $t$, as $t\to\infty$). In particular, 
\be\label{eq:NOSind}
\big\| j_{\m, x}^{R}(t)  - j_{\m, x}(t) \big\| \leq \e(R)\,,\quad \text{for some $\e(R)\to 0$ as $R\to\infty$}.
\ee
Therefore, this bound implies that, for $\widetilde\e(R) \leq c \e(R)$, using the boundedness of the fermionic operators:
\be
\big|\big\langle \big(j_{\m, x}(t) - j_{\m, x}^{R}(t)\big) j_{\n, 0} \big\rangle_{\beta,L}\big| \leq \| j_{\m, x}(t) - j_{\m, x}^{R}(t) \|\| j_{\n, 0} \| \leq \widetilde\e(R)\;.
\ee
Consider now the first term in the right-hand side of Eq. (\ref{eq:decomp}). We claim that the $\beta, L\to \infty$ limit exists, for fixed $R$. To prove this we use that, since $\| \mathcal{H}^{R}_{L} \|\leq CR$,
\be\label{eq:RR}
j_{\m, x}^{R}(t)  = \sum_{n\geq 0} \frac{t^{n}}{n!}\text{ad}^{n}_{\mathcal{H}_{L}^{R}}(j_{\m, x})\;,\qquad \| \text{ad}^{n}_{\mathcal{H}_{L}^{R}}(j_{\m, x}) \|\leq \|j_{\m, x}\|(2C)^{n}R^{n}\;, 
\ee
where $\text{ad}^{n}_{\mathcal{H}_{L}^{R}}(j_{\m, x})$ is the $n$-fold commutator of $j_{\m, x}$ with $\mathcal{H}_{L}^{R}$, and $C$ is a constant independent of $L, R$. We then write $\big\langle j_{\m, x}^{R}(t) j_{\n, 0} \big\rangle_{\beta,L} = \sum_{n\geq 0} \frac{t^{n}}{n!} \big\langle \text{ad}^{n}_{\mathcal{H}_{L}^{R}}(j_{\m, x}) j_{\n, 0} \big\rangle_{\b, L}$; being the sum convergent uniformly in $\beta, L$, thanks to Eq. (\ref{eq:RR}), to prove the existence of the limit $\big\langle j_{\m, x}^{R}(t) j_{\n, 0} \big\rangle_{\infty} = \lim_{\beta,L\to\infty}\big\langle j_{\m, x}^{R} (t) j_{\n, 0} \big\rangle_{\beta,L}$ it is enough to prove the existence of $\lim_{\beta,L\to\infty} \big\langle \text{ad}^{n}_{\mathcal{H}_{L}^{R}}(j_{\m, x}) j_{\n, 0} \big\rangle_{\beta,L} $ for all $n$. This is the zero temperature, infinite volume limit of a {\it static} correlation function in configuration space, which exists thanks to item $(i)$. Therefore,
\bea
-\widetilde \e(R) &\leq&  \liminf_{\beta,L\to\infty} \big\langle j_{\m, x}(t) j_{\n, 0} \big\rangle_{\beta,L} - \lim_{\b, L\to \infty}\big\langle j_{\m, x}^{R} j_{\n, 0} \big\rangle_{\b, L} \\&\leq&   \limsup_{\beta,L\to\infty}  \big\langle j_{\m, x}(t) j_{\n, 0} \big\rangle_{\beta,L} - \lim_{\b, L\to \infty}\big\langle j_{\m, x}^{R}  j_{\n, 0} \big\rangle_{\b, L} \le \widetilde \e(R) \,,\nn
\eea
that is:
\be
  \limsup_{\beta,L\to\infty} \big\langle j_{\m, x}(t) j_{\n, 0} \big\rangle_{\beta,L} -\widetilde \e(R)\le  \lim_{\b, L\to \infty}\big\langle j_{\m, x}^{R}(t) j_{\n, 0} \big\rangle_{\b, L}
\le \liminf_{\beta,L\to\infty} \big\langle j_{\m, x}(t) j_{\n, 0} \big\rangle_{\beta,L}+ \widetilde \e(R).\nn
\ee
Letting $R\to\infty$, we find that the liminf and limsup coincide, as desired. This proves Eq. (\ref{eq:bd}), and concludes the proof of Proposition \ref{prp:lim}.
\end{proof}

\section{Conclusions}\label{sec:concl}

We used rigorous RG methods to study the ground-state transport properties of non-integrable quantum spin chains. If combined with complex analytic ideas, these methods give access {\it real-time} transport coefficients. In particular, we proved the identity between Euclidean and canonical Drude weights and susceptibilities. As far as we know, this is the first rigorous result about the real-time transport coefficients of a non-integrable quantum spin chain, in the thermodynamic limit. Moreover, as a corollary of previous results \cite{BM, BM1, BM2, BM3}, our theorem proves the validity of the Haldane scaling relation for the real-time transport coefficients.

As a result, the qualitative properties of the zero temperature canonical Drude weight are unaffected by the presence of integrability breaking terms: for small perturbations, the Drude weight of the quantum spin chain stays nonzero. It is a very interesting open problem to extend these methods to the {\it positive temperature} regime: there, the Drude weight is expected to be nonzero or zero depending on whether the model is integrable or not, see {\it e.g.} \cite{Z1, Z2, IP1, IP2, ASP}. The discussion of the proof of our main result, Theorem \ref{thm:wick}, locates the problem in going to positive temperatures in the approximation of the adiabatic parameter $\eta$ with its Matsubara counterpart $\eta_{\beta}\in \frac{2\pi}{\beta}\mathbb{Z}$. We believe that a refinement of the above strategy could shed light on this challenging open problem.
\medskip

\noindent{\bf Acknowledgements.} V. M. has received funding from the European Research Council (ERC)
under the European Union's Horizon 2020 research and innovation programme (ERC CoG UniCoSM, grant agreement n.724939). The work of M. P. has been supported by the Swiss National Science Foundation via the grant ``Mathematical Aspects of Many-Body Quantum Systems''. We thank K. Gawedzki and H. Spohn for valuable discussions, and for comments on an earlier version of the manuscript. In particular, we thank H. Spohn for a stimulating discussion on the equivalence of thermal and canonical Drude weights, that motivated Appendix \ref{app:B}  below.

\appendix
\section{Formal derivation of Kubo formula}\label{app:A}
In this appendix we give formal derivation of Kubo formula for the conductivity of the quantum spin chain, Eq. (\ref{ss}). Eq. (\ref{ss}) provides the linear response coefficient that allows to describe the variation of the average current density after introducing a weak external electric field, assumed to be uniform in space. The electric field is adiabatically switched on at $t=-\infty$, starting from the thermal state $\r_{\b}$ of the Hamiltonian $\mathcal{H}$.

In a rigorous derivation of Kubo formula, one has to take the thermodynamic limit {\it before} the limit of vanishing perturbation. Controlling the thermodynamic limit for a fixed external perturbation poses a technical challenge, that for interacting systems has only been solved for {\it gapped} systems, \cite{Ba, Mo, T}. Instead, the models we are considering are {\it gapless}: the problem of deriving Kubo formula for this class of interacting quantum spin chains is wide open. We shall not try to study this interesting question here; instead, we will focus on the linear response of the current operator, formally neglecting higher order terms.

For simplicity, let us directly consider the case $L=\infty$: $\L_{L} = \mathbb{Z}$. Let $\mathcal{X} = \sum_{x} x a^{+}_{x}a^{-}_{x}$ be the second quantization of the position operator. Consider the time-dependent Hamiltonian $\mathcal{H}(t) = \mathcal{H} + e^{\eta t} E \mathcal{X}$, for $t\in (-\infty, 0]$ and $\eta >0$, $E\in \mathbb{R}$. Let $\rho(t)$ denote the solution of the Schr\"odinger-von Neumann equation:
\be
i\partial_{t} \r(t) = [\mathcal{H}(t), \rho(t)]\;,\qquad \r(-\infty) = \r_{\beta}\;.
\ee
We are interested in a formal expansion in $E$ for the average current, in the limit $\eta \to 0^{+}$. An application of Duhamel formula gives:
\bea\label{eq:r1}
\Tr\, j_{x} \r(t) &=& \Tr\, j_{x} \r_{\beta} - i\int_{-\infty}^{0} dt\, e^{\eta t} \Tr\, j_{x} [ E \mathcal{X}(t), \rho_{\beta} ] + o(E)\nn\\
&=& i\int_{-\infty}^{0} dt\, e^{\eta t} \Tr [ j_{x}(-t), E\mathcal{X} ]\rho_{\beta} + o(E)\;,
\eea
since $\Tr\, j_{x} \r_{\beta} = 0$. Then,
\bea\label{eq:JX}
[ j_{x}(-t), \mathcal{X} ] &=& [ e^{-i \mathcal{H} t} j_{x} e^{i \mathcal{H} t}, \mathcal{X} ]\nn\\ &=& e^{-i \mathcal{H}t} [ j_x , \mathcal{X}] e^{i\mathcal{H}t} + [ e^{-i\mathcal{H}t}, \mathcal{X} ] j_x e^{i \mathcal{H}t}  + e^{-i\mathcal{H}t} j_x [e^{i\mathcal{H}t}, \mathcal{X}]\;,\nn\\
\eea
where, setting $\mathcal{J} \equiv \sum_{x} j_{x}$:
\bea
[ e^{-i\mathcal{H}t}, \mathcal{X} ] &=& e^{-i \mathcal{H} t} \mathcal{X} - \mathcal{X} e^{-i\mathcal{H} t} = \int_{t}^{0} ds\, \frac{d}{ds} e^{-i\mathcal{H}(t - s)} \mathcal{X} e^{-i\mathcal{H}s}\nn\\
&=& i\int_{t}^{0} ds\, e^{-i \mathcal{H}(t - s)} [ \mathcal{H}, \mathcal{X} ]  e^{-i\mathcal{H}s}\nn\\&=& \int_{t}^{0}ds\, e^{-i\mathcal{H}t} \mathcal{J}(s) \equiv -\int_{t}^{0}ds\,  \mathcal{J}(s) e^{i\mathcal{H}t}
\eea
and:
\be\label{eq:JxX}
[ j_x , \mathcal{X}] = -\frac{it}{2} [a^{+}_{x+1} a^{-}_{x} + a^{+}_{x}a^{-}_{x+1}] \equiv i \D_{x}\;.
\ee
Therefore, plugging (\ref{eq:JX})-(\ref{eq:JxX}) into Eq. (\ref{eq:r1}):
\bea\label{eq:s}
i\int_{-\infty}^{0} dt\, e^{\eta t} \Tr [ j_{x}(-t), \mathcal{X} ]\rho_{\beta} &=& -\int_{-\infty}^{0} dt\, e^{\eta t} \Tr\, \D_{x} \rho_{\beta} + i \int_{-\infty}^{0} dt\, e^{\eta t} \int_{t}^{0} ds\, \Tr\, [ \mathcal{J}(s), j_{x} ] \rho_{\beta} \nn\\
&=& -\frac{1}{\eta} \Tr\, \D_{x} \r_{\b} +  i \int_{-\infty}^{0} ds\, \int_{-\infty}^{s} dt\, e^{\eta t}  \Tr\, [ \mathcal{J}(s), j_{x} ] \rho_{\beta}\nn\\
&=& \frac{1}{\eta} \Big[ -\Tr\, \D_{x} \r_{\b} + i\int_{-\infty}^{0} ds\, e^{\eta s}  \Tr\, [ \mathcal{J}(s), j_{x} ] \rho_{\beta}  \Big]\;.
\eea
By translation invariance, the above expression does not depend on $x$. Thus, the right-hand side of Eq. (\ref{eq:s}) reproduces Eq. (\ref{ss}), for $p = 0$ and $\eta \to 0^{+}$ (notice that $\hat j_{0}\equiv \mathcal{J}$). In general, $p\neq 0$ allows to take into account a space modulation of the external field.

\section{On the equivalence between thermal and canonical Drude weight}\label{app:B}
In this appendix we discuss the equivalence between a suitably regularized version of the thermal Drude weight, and the canonical Drude weight. Given two operators $A, B$, we define their {\it Kubo scalar product} as:
\be
\langle A B\rangle^{K}_{\b,L} := \int_{0}^{\beta} dx_{0}\, \langle A(-ix_{0}) B \rangle_{\b,L}\;.
\ee
We then notice that:
\be
i\partial_{t} \langle A(t) B \rangle^{K}_{\b,L} = \langle [ A(t), B ] \rangle^{K}_{\b,L}\;.
\ee
Therefore, one formally has:
\bea\label{eq:tildeD}
\widetilde D^{\text{(Th)}} _\b
= \lim_{t\to\infty} \lim_{L\to \infty} \frac{1}{L}\langle \mathcal{J}(t) \mathcal{J} \rangle^{K}_{\b, L} &=& \lim_{L\to \infty} \frac{1}{L}\langle \mathcal{J} \mathcal{J}\rangle_{\b, L}^{K} - i \int_{0}^{\infty} ds\, \lim_{L\to \infty} \frac{1}{L}\langle [ \mathcal{J}(s), \mathcal{J} ] \rangle_{\b,L} \nn\\ &\equiv&\lim_{L\to \infty} \frac{1}{L}\langle \mathcal{J} \mathcal{J}\rangle_{\b, L}^{K} + i \int_{-\infty}^{0} ds\, \lim_{L\to \infty} \frac{1}{L} \langle [ \mathcal{J}(s), \mathcal{J} ] \rangle_{\b, L}\;.\nn
\eea
The main problem in making sense of the above identity is the existence of the time integration: for nonintegrable systems, proving that the integral converges is a very hard open problem. Therefore, let us introduce the following regularized version of the thermal Drude weight, and positive and zero temperature:
\bea
\widehat D^{\text{(Th)}} _\b &=& \lim_{L\to \infty} \frac{1}{L}\langle \mathcal{J} \mathcal{J}\rangle^{K}_{\b, L} + \lim_{\eta\to 0^{+}}i \int_{-\infty}^{0} ds\, e^{\eta s} \lim_{L\to \infty} \frac{1}{L} \langle [ \mathcal{J}(s), \mathcal{J} ] \rangle_{\b,L}\nn\\
\widehat D^{\text{(Th)}} _\infty &=& \lim_{\b, L\to \infty}\frac{1}{L} \langle \mathcal{J} \mathcal{J}\rangle^{K}_{\b, L} + \lim_{\eta\to 0^{+}}i \int_{-\infty}^{0} ds\, e^{\eta s} \lim_{\b, L\to \infty} \frac{1}{L}\langle [ \mathcal{J}(s), \mathcal{J} ] \rangle_{\b,L}\;.
\eea
Of course, $\widehat{D}^{\text{(Th)}} = \widetilde{D}^{\text{(Th)}}$, if the real-time correlations decay fast enough. Such time regularization plays the same role of the adiabatic factor in the derivation of Kubo formula, see Appendix \ref{app:A}. This regularized version of the thermal Drude weight matches the canonical formulation, Eqs. (\ref{eq:D}), {\it provided} one can show that 
\be\label{eq:Deq}
\lim_{L\to \infty} \frac{1}{L}\langle \mathcal{J} \mathcal{J} \rangle^{K}_{\b, L} = - \lim_{L\to \infty} \langle \D_{x} \rangle_{\b,L}\;.
\ee
To prove Eq. (\ref{eq:Deq}), we proceed as follows. Consider the lattice continuity equation:
\be\label{eq:cons}
\partial_{t} \r_{x}(t) + d_{x} j_{x}(t) = 0\;,
\ee
with $d_{x} j_{x} = j_{x} - j_{x-1}$ the discrete lattice derivative. In imaginary times, Eq. (\ref{eq:cons}) reads:
\be\label{eq:consI}
i\partial_{x_{0}} \r_{x}(-ix_{0}) + \partial_{x} j_{x}(-ix_{0}) = 0\;.
\ee
Eq. (\ref{eq:consI}) can be used to derive Ward identities for the current-current correlations, as follows. Let $j_{0,x} \equiv \r_{x}$, $j_{1,x} \equiv j_{x}$. Recall:
\bea
\langle {\bf T}\, j_{0, x}(-ix_{0})\,; j_{\n, 0}(-iy_{0}) \rangle_{\b,L} &=& \theta(x_{0} - y_{0}) \langle  j_{0, x}(-ix_{0})\,; j_{\n, 0}(-iy_{0}) \rangle_{\b,L}\nn\\&& + \theta(y_{0} - x_{0}) \langle j_{\n, 0}(-iy_{0}) j_{0, x}(-ix_{0}) \rangle_{\b,L}\;;
\eea
therefore, using the continuity equation Eq. (\ref{eq:consI}):
\bea
i\partial_{x_{0}}\langle {\bf T}\, j_{0, x}(-ix_{0})\,; j_{\n, 0}(-iy_{0}) \rangle_{\b,L} &=& \langle {\bf T}\,  i\partial_{x_{0}}j_{0, x}(-ix_{0})\,; j_{\n, 0}(-iy_{0}) \rangle_{\b,L}\nn\\&& \qquad + i\langle [ j_{0, x}\, , j_{\n, 0} ] \rangle_{\b,L} \d(x_{0} - y_{0})\nn\\
&\equiv& -\langle {\bf T}\,  \partial_{x}j_{1, x}(-ix_{0})\,; j_{\n, 0}(-iy_{0}) \rangle_{\b,L}\nn\\&& \qquad + i\langle [ j_{0, x}\, , j_{\n, 0} ] \rangle_{\b,L} \d(x_{0} - y_{0})\;.
\eea
Now, consider the Fourier transform:
\be\label{eq:JJ}
\langle {\bf T}\, \hat j_{\m,\pp}\,; \hat j_{\n,-\pp} \rangle_{\beta, L} \equiv \frac{1}{\beta} \int_{0}^{\beta} d x_{0} \int_{0}^{\beta} d y_{0}\,e^{-ip_{0}(x_{0} - y_{0})} \sum_{x = -L/2}^{L/2} e^{-ipx} \langle {\bf T}\, j_{\m, x}(-i x_{0})\,; j_{\n, 0}(-i y_{0}) \rangle_{\b, L}\;,
\ee
with ${\bf T}$ the fermionic time ordering, and $\pp = (p_{0}, p) \in \frac{2\pi}{\beta} \mathbb{Z} \times \frac{2\pi}{L} \mathbb{Z}$. In Eq. (\ref{eq:JJ}), the $x_{0}, y_{0}$ integrations have to be understood as integrals over circles of length $\beta$ (the imaginary time evolution of the current operators is extended periodically outside of $[0, \beta)$). We have, integrating by parts the time-derivative:
\bea
p_{0}\langle {\bf T}\, \hat j_{0,\pp}\,; \hat j_{\n,-\pp} \rangle_{\beta,L} &=& \frac{1}{\beta} \int_{0}^{\beta} d x_{0} \int_{0}^{\beta} d y_{0}\, p_{0}e^{-ip_{0}(x_{0} - y_{0})}\sum_{x} e^{-ipx} \langle {\bf T}\, j_{0, x}(-i x_{0})\,; j_{\n, 0}(-iy_{0}) \rangle_{\b,L}\nn\\
&=&  \frac{1}{\beta} \int_{0}^{\beta} d x_{0}\,  \int_{0}^{\beta} d y_{0}\, (i\partial_{x_{0}} e^{-ip_{0}(x_{0} - y_{0})})\sum_{x} e^{-ipx} \langle {\bf T}\, j_{0, x}(-ix_{0})\,; j_{\n, 0}(-i y_{0}) \rangle_{\b,L}\nn\\
&=& -\frac{1}{\beta} \int_{0}^{\beta} dx_{0} \int_{0}^{\beta} dy_{0}\, e^{-ip_{0}(x_{0} - y_{0})} \sum_{x} e^{-ipx} i\partial_{x_{0}}\langle {\bf T}\, j_{0, x}(-i x_{0})\,; j_{\n, 0}(-i y_{0}) \rangle_{\b,L}\;.\nn
\eea
Then, thanks to the continuity equation (\ref{eq:cons}), we get:
\be\label{eq:WIf}
p_{0}\langle {\bf T}\, \hat j_{0,\pp}\,; \hat j_{\n,-\pp} \rangle_{\beta,L} = \eta(p) \langle {\bf T}\, \hat j_{1,\pp}\,; \hat j_{\n, -\pp} \rangle_{\b,L} - i\sum_{x} e^{-ipx} \langle [ \hat j_{0, x}\, , \hat j_{\n, 0} ] \rangle_{\b,L}
\ee
with $\eta(p) = (1 - e^{-ip})$. Let $\n = 1$, and set $p_{0} = 0$, $p_{1}\neq 0$. Eq. (\ref{eq:WIf}) implies:
\be\label{eq:JJ1}
\eta(p) \langle {\bf T}\, \hat j_{1,(0,p)}\,; \hat j_{1, (0,-p)} \rangle_{\beta,L} = i\sum_{x} e^{-ipx} \langle [ j_{0, x}\, , j_{1, 0} ] \rangle_{\beta,L} = \frac{t}{2}(1 - e^{-ip}) \langle a^{+}_{1}a^-_{0} + a^{+}_{0}a^-_{1}\rangle_{\beta,L}\;.
\ee
Now, notice that $\langle {\bf T}\, \hat j_{1,\V0}\,; \hat j_{1,\V0} \rangle_{\b,L} = L^{-1}\langle \mathcal{J} \mathcal{J}\rangle^{K}_{\b,L}$. Hence, from Eq. (\ref{eq:JJ1}), recalling Eq. (\ref{eq:JxX}), and setting $\langle \cdot \rangle_{\b} \equiv \lim_{L\to \infty} \langle \cdot \rangle_{\b,L}$:
\be
\lim_{L\to \infty} \frac{1}{L}\langle \mathcal{J} \mathcal{J}\rangle^{K}_{\b,L}  = \lim_{p\to 0}\langle {\bf T}\, \hat j_{1,(0,p)}\,; \hat j_{1, (0,-p)} \rangle_{\beta} =  \frac{t}{2}\langle a^{+}_{1}a_{0} + a^{+}_{0}a_{1}\rangle_{\beta} \equiv -\langle \D_{x} \rangle_{\b}\;,
\ee
which proves Eq. (\ref{eq:Deq}).

\end{document}